\documentclass[twocolumn,showpacs,prb]{revtex4-1}%
\usepackage{amsfonts}
\usepackage{amsmath}
\usepackage{amssymb}
\usepackage{subfigure}
\usepackage{graphicx}
\usepackage{txfonts}
\usepackage{epstopdf}
\usepackage{makecell}
\usepackage[mathlines]{lineno}
\usepackage[colorlinks,linkcolor=blue,citecolor=blue,urlcolor=blue]{hyperref}%
\providecommand{\U}[1]{\protect\rule{.1in}{.1in}}

\begin{document}

\title{Geometric density of states of electronic structures for local responses: Phase information from the amplitudes of STM measurement}
\author{Shu-Hui Zhang$^{1}$}
\email{shuhuizhang@mail.buct.edu.cn}
\author{Jin Yang$^{2}$}
\author{Ding-Fu Shao$^{3}$}
\author{Jia-Ji Zhu$^{4}$}
\author{Wen Yang$^{2}$}
\email{wenyang@csrc.ac.cn}
\author{Kai Chang$^{5}$}
\email{kchang@semi.ac.cn}

\affiliation{$^{1}$College of Mathematics and Physics, Beijing University of Chemical Technology, Beijing,
100029, China}
\affiliation{$^{2}$Beijing Computational Science Research Center, Beijing 100193, China}
\affiliation{$^{3}$Key Laboratory of Materials Physics, Institute of Solid State Physics, HFIPS, Chinese Academy of Sciences,
Hefei 230031, China}
\affiliation{$^{4}$School of Science, Chongqing University of Posts and Telecommunications, Chongqing 400061, China}
\affiliation{$^{5}$SKLSM, Institute of Semiconductors, Chinese Academy of Sciences, Beijing 100083, China}

\begin{abstract}
Electronic band structures underlie the physical properties of crystalline
materials, their geometrical exploration renovates the conventional cognition and
brings about novel applications. Inspired by geometry phases, we introduce a geometric amplitude named
as the geometric density of states (GDOS) dictated by the
differential curvature of the constant-energy contour. The GDOS determines the amplitude of the real-space Green's function making it attain the ultimate
expression with transparent physics. The local responses of crystalline
materials are usually formulated by the real-space Green's function, so the
relevant physics should be refreshed by GDOS. As an example of
local responses, we suggest using scanning tunneling microscopy (STM) to
characterize the surface states of three-dimensional topological insulator
under an in-plane magnetic field. The GDOS favors the straightforward
simulation of STM measurement without resorting to Fourier transform of
the real-space measurement, and also excavates the unexplored potential of STM measurement to extract the phase information of wavefunction through its amplitude, i.e., the
spin and curvature textures. Therefore, the proposed GDOS deepens the
understanding of electronic band structures and is indispensable in local
responses, and it should be universal for any periodic systems.

\end{abstract}
\maketitle

\affiliation{$^{1}$College of Mathematics and Physics, Beijing University of Chemical Technology, Beijing,
100029, China}
\affiliation{$^{2}$Beijing Computational Science Research Center, Beijing 100193, China}
\affiliation{$^{3}$Key Laboratory of Materials Physics, Institute of Solid State Physics, HFIPS, Chinese Academy of Sciences,
Hefei 230031, China}
\affiliation{$^{4}$School of Science, Chongqing University of Posts and Telecommunications, Chongqing 400061, China}
\affiliation{$^{5}$SKLSM, Institute of Semiconductors, Chinese Academy of Sciences, Beijing 100083, China}

\affiliation{$^{1}$College of Mathematics and Physics, Beijing University of Chemical Technology, Beijing,
100029, China}
\affiliation{$^{2}$Beijing Computational Science Research Center, Beijing 100193, China}
\affiliation{$^{3}$Key Laboratory of Materials Physics, Institute of Solid State Physics, HFIPS, Chinese Academy of Sciences,
Hefei 230031, China}

\affiliation{$^{1}$College of Mathematics and Physics, Beijing University of Chemical Technology, Beijing,
100029, China}
\affiliation{$^{2}$Beijing Computational Science Research Center, Beijing 100193, China}
\affiliation{$^{3}$Key Laboratory of Materials Physics, Institute of Solid State Physics, HFIPS, Chinese Academy of Sciences,
Hefei 230031, China}

\textit{Introduction}.---As a successful application of quantum mechanics, the
electronic band theory was developed to describe the motion of electrons or
quasiparticles in crystalline materials with periodic structures in real
space. The conventional electronic band theory describes the relation of
energies and momenta of quasiparticles, i.e., the electronic band structure
underlying the mechanical, thermal, optical and electrical properties of
crystalline materials\cite{Kittel2005}. The profound insights into the band
structure flourish into the modern band theory with geometrical phase\cite{rspa.1984.0023,s42254-019-0071-1} as an outstanding example. Geometrical phase embodies the geometrical
information of band structure and is responsible for Berry physics and various
novel phenomena\cite{RevModPhys.82.1959}. At present, geometrical phase has become an
essential ingredient of band structure, this inspires us to discover the
analogous concept.

The macroscopic and global measurement, e.g., various Hall transport
experiments, is usually used to reveal Berry physics since Berry curvature
intrinsically provides an overall perspective of the electronic band
structure\cite{vanderbilt2018}. Conversely, the local measurement is expected
to reflect the local information of crystalline materials, which is strongly
required in the field of low-dimensional
physics\cite{s42254-021-00293-7,ZhangJPCM2022}. Recently, two experiments use
scanning tunneling microscopy (STM) to measure the Friedel oscillation (FO)
induced by intentional introduced single impurity, then count the wavefront
dislocations with the specific number corresponding to the Berry phase of
monolayer and bilayer graphene\cite{s41586-019-1613-5,PhysRevLett.125.116804}.
Thus, these two experiments also demonstrate the possibility to extract the
global Berry phase\cite{CRPHYS2021}. Hence, comparing to the global
measurement, the local measurement provides information with higher complexity
helpful to build the physical understanding of the novel materials. Furthermore, there are emergent
nonmagnetic, magnetic, topological two-dimensional (2D) materials and exotic surface states of
three-dimensional topological insulators and semimetals, their novel physics
and potential
applications\cite{RevModPhys.81.109,HanNatureNanotechnology2014,s11467-018-0859-y,natrevmats.2016.42,s41586-019-1573-9,aac9439,RevModPhys.92.021003}%
promote the ever-increasing demand of local information. 

\begin{figure}[ptbh]
\includegraphics[width=0.9\columnwidth,clip]{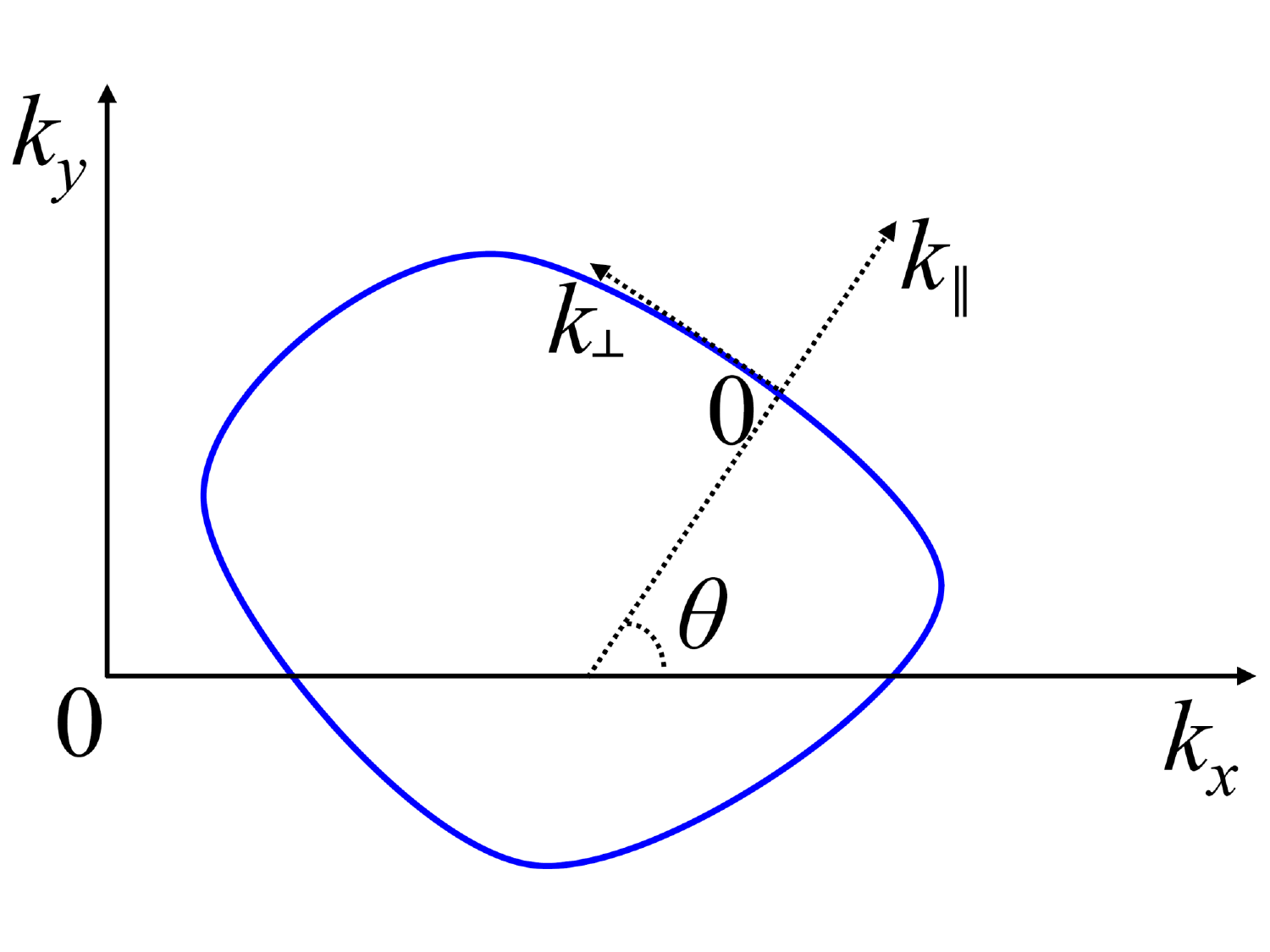}\caption{Two coordinate
systems and their relation. Choosing an energy $\varepsilon$, a general
constant-energy contour is given through the energy dispersion, i.e.,
$E(\mathbf{k})=\varepsilon$, for the two-dimensional quasiparticles with the
momentum $\mathbf{k=(}k_{x},k_{y}\mathbf{)}$ in the coordinate system
$k_{x}-k_{y}$. The gradient of constant-energy contour or the group velocity
$\mathbf{v}=\nabla E(\mathbf{k})$ helps define the unit vectors $\mathbf{n}%
_{\mathbf{\parallel}}=\mathbf{v}\left\vert \mathbf{v}\right\vert $ and
$\mathbf{n}_{\mathbf{\perp}}=\mathbf{z}\times\mathbf{v/}\left\vert
\mathbf{v}\right\vert $ with $\mathbf{z}$ pointing outside of the plane. And
$\theta$ is the azimuthal angle of $\mathbf{v}$. Thus, a local coordinate
system $k_{\parallel}-k_{\perp}$ is constructed, in which $k_{\parallel
}=\mathbf{k\cdot n}_{\mathbf{\parallel}}$ and $k_{\perp}=\mathbf{k\cdot
n}_{\mathbf{\perp}}$.}%
\label{GCEC}%
\end{figure}

In this study, we provide a local perspective of the electronic band structure
by introducing microscopic description on the density of states (DOS). The
conventional DOS provides a shortcut to catch the useful information from the
electronic structure, which may be the most important concept for
understanding physical properties of crystalline materials and is formulated
in undergraduate textbooks\cite{1-s2.0-S277294942200002X-main}. The
microscopic DOS is dictated by the differential curvature of the
constant-energy contour (CEC) of band structure, then is called geometric
DOS (GDOS), which represents one kind of geometrical amplitude in contrast to geometrical phase\cite{RevModPhys.82.1959}. The GDOS governs the amplitude of the real-space Green's function (GF) and makes
it attain the ultimate expression with transparent physics. Since the local
responses of crystalline materials are formulated by the real-space GF, the
GDOS concept must refresh the understanding of previous local physical
properties\cite{SimonJPDAP2011,Bena2016,ChenJPCM2017,adma.201707628,ZhengAIPX2018,LinJPCM2020,s42254-021-00293-7,ZhangJPCM2022,RevModPhys.82.1633,cryst3010049,SettnesPRL2014}%
. As an application for simulating STM experiment, we explore the possible
modification of the electronic band structure of the surface states of
three-dimensional topological insulator by an in-plane magnetic
field\cite{PhysRevB.101.041408,PhysRevB.103.155415,PhysRevLett.127.076601}, which is blind for angle-resolved photoemission spectroscopy\cite{s42254-021-00293-7}.
The GDOS, through the amplitude of real-space GF, provides a novel protocol to
understand the STM measurement. More importantly, the GDOS dictates the STM
detection of phase information, namely spin and curvature textures, through the amplitudes. Therefore, the proposed GDOS is a
pivotal concept for the electronic structure and useful for the local responses,
and should also be universal in any periodic systems.

\textit{Definition of geometric density of states}.---Fig. \ref{GCEC} shows a
general CEC $E(\mathbf{k})=\varepsilon$ for the 2D
quasiparticles with the momentum $\mathbf{k=(}k_{x},k_{y}\mathbf{)}$ in the
coordinate system $x-y$. The gradient of the CEC is the group velocity
$\mathbf{v}=\nabla_{\mathbf{k}}E(\mathbf{k})$. As shown by Fig. \ref{GCEC}, a
local coordinate system $k_{\parallel}-k_{\perp}$ is constructed through
$k_{\parallel}=\mathbf{k\cdot n}_{\mathbf{\parallel}}$ and $k_{\perp
}=\mathbf{k\cdot n}_{\mathbf{\perp}}$\ by using $\mathbf{n}_{\mathbf{\parallel
}}=\mathbf{v/}\left\vert \mathbf{v}\right\vert $ and $\mathbf{n}%
_{\mathbf{\perp}}=\mathbf{z}\times\mathbf{v/}\left\vert \mathbf{v}\right\vert
$. The conventional DOS is defined as $\rho_{0}(\varepsilon)\equiv\lim\delta
N/\delta\varepsilon$\cite{Kittel2005}, in which $\delta N$ represents the
number of states in the energy range $\delta\varepsilon$. \ In analogy to
$\rho_{0}(\varepsilon)$, we introduce the microscopic DOS $\rho(\varepsilon
,\theta)\equiv\lim\delta N/(\delta\varepsilon\delta\theta)$ with $\theta$ the
azimuthal angle of the group velocity in the coordinate system $x-y$. From
Fig. \ref{GCEC}, we arrive at%
\begin{equation}
\rho(\varepsilon,\theta)=\frac{1}{4\pi^{2}}\frac{\delta k_{\parallel}\delta
s}{\delta\varepsilon\delta\theta}=\frac{1}{4\pi^{2}v\kappa}, \label{DDOS}%
\end{equation}
due to the group velocity $v\mathbf{=\lim}\delta\varepsilon/\delta
k_{\parallel}$ and the curvature of the CEC:
\begin{equation}
\kappa=\lim\delta\theta/\delta s, \label{DC}%
\end{equation}
with $\delta s$ the curve length of the CEC. Since $\rho(\varepsilon,\theta)$
is determined by the differential curvature $\kappa$, so it attains the name
GDOS. There are two implications of Eq. (\ref{DDOS}). On one hand, it
gives the simple calculation of the GDOS since $\kappa$ is a well-known
mathematical quantity\cite{Carmo2016,GDOS2023}. On the other hand, it allows the
measurement of $\kappa$ in experiment as shown below, which is inherently interesting.

\textit{Real-space Green's function}.---The key of local responses is the
real-space GF\cite{Economou1983}. To consider a 2D system with the Hamiltonian
in the diagonal basis:
\begin{equation}
\mathcal{H}\mathbf{(k)=}\sum_{\lambda}E_{\lambda}(\mathbf{k})\left\vert
u_{\lambda}(\mathbf{k})\right\rangle \left\langle u_{\lambda}(\mathbf{k}%
)\right\vert .
\end{equation}
Here, $\lambda$ denotes the band index, $\mathbf{k=(}k_{x},k_{y}\mathbf{)}$ is the
momentum in the coordinate system $x-y$, $E_{\lambda}(\mathbf{k})$ and
$\left\vert u_{\lambda}(\mathbf{k})\right\rangle $ are the eigenvalues and
eigenstates, respectively. The corresponding real-space GF is $\mathcal{G}%
(\varepsilon,\mathbf{r})=\sum_{\lambda}\mathbf{g}_{\lambda}(\varepsilon
,\mathbf{r})$, which is the sum of contributions from different energy bands:%
\begin{equation}
\mathbf{g}_{\lambda}(\varepsilon,\mathbf{r})=%
{\displaystyle\iint}
d\mathbf{k}\frac{e^{i\mathbf{k\cdot r}}}{4\pi^{2}}\frac{\left\vert u_{\lambda
}(\mathbf{k})\right\rangle \left\langle u_{\lambda}(\mathbf{k})\right\vert
}{(\varepsilon+i0^{+})-E_{\lambda}(\mathbf{k})}.
\end{equation}
To decompose $\mathbf{k\equiv}(k_{\mathbf{\parallel}}\mathbf{,}k_{\perp})$
into one parallel component $k_{\parallel}$\ and one perpendicular component
$k_{\perp}$ with respect to $\mathbf{r}$ (cf. Fig. \ref{GCEC}), thus%
\begin{equation}
\mathbf{g}_{\lambda}(\varepsilon,\mathbf{r})=\int\frac{dk_{\perp}}{2\pi
}\left(  \int\frac{dk_{\parallel}}{2\pi}e^{i\mathbf{k\cdot r}}\frac{\left\vert
u_{\lambda}(\mathbf{k})\right\rangle \left\langle u_{\lambda}(\mathbf{k}%
)\right\vert }{(\varepsilon+i0^{+})-E_{\lambda}(k_{\mathbf{\parallel}%
}\mathbf{,}k_{\perp})}\right)  .
\end{equation}
Usually, the band index $\lambda$ can be omitted by assuming that there is
only one on-shell solution $k_{\mathbf{\parallel,}+}(k_{\perp})$ for
$E_{\lambda}(k_{\mathbf{\parallel}}\mathbf{,}k_{\perp})=\varepsilon$ to ensure
the group velocity $v_{\mathbf{\parallel}}(k_{\perp})\equiv\partial
_{k_{\parallel}}E(\mathbf{k})|_{k_{\mathbf{\parallel}}=k_{\mathbf{\parallel
,}+}(k_{\perp})}$ along $\mathbf{r}$ with assuming its component $x>0$, i.e.,
$v_{\mathbf{\parallel}}(k_{\perp})$ is positive or equivalently
$\operatorname{Im}k_{\mathbf{\parallel,}+}>0$. To perform the Taylor expansion
about $k_{\mathbf{\parallel}}$ for $E(\mathbf{k})$ near $k_{\mathbf{\parallel
,}+}(k_{\perp})$, i.e., $E(k_{\mathbf{\parallel}},k_{\perp})\approx
\varepsilon+v_{\mathbf{\parallel}}(k_{\perp})\left[  k_{\mathbf{\parallel}%
}-k_{\mathbf{\parallel,}+}(k_{\perp})\right]  $, which helps finish the
contour integral of $\mathbf{g}(\varepsilon,\mathbf{r})$ over
$k_{\mathbf{\parallel}}$ in the upper half complex plane:%
\begin{equation}
\mathbf{g}(\varepsilon,\mathbf{r})\approx\int\frac{dk_{\perp}}{2\pi}%
\frac{e^{ik_{\mathbf{\parallel,}+}(k_{\perp})r}}{iv_{\mathbf{\parallel}}%
}\left\vert u(k_{\mathbf{\parallel,}+},k_{\mathbf{\perp}})\right\rangle
\left\langle u(k_{\mathbf{\parallel,}+},k_{\mathbf{\perp}})\right\vert .
\label{GR1}%
\end{equation}
Therefore, Eq. (\ref{GR1}) implies that all right-going on-shell eigenstates
(either traveling or evanescent) contribute to GF. In the stationary phase
approximation\cite{PhysRev.149.519,PhysRevB.83.035427,PhysRevB.85.125314} for
large $r$, the contribution to Eq. (\ref{GR1}) should be dominated by
$k_{\mathbf{\perp}}$ near the saddle point $k_{\mathbf{\perp,}c}$ determined
by solution of $\partial_{k_{\mathbf{\perp}}}k_{\mathbf{\parallel}%
}(k_{\mathbf{\perp}})=0$. So the classical momenta can be defined, i.e.,
$\mathbf{k}_{c}\equiv(k_{\mathbf{\parallel,}c},k_{\mathbf{\perp,}c})$ with
$k_{\mathbf{\parallel,}c}\equiv k_{\mathbf{\parallel,}+}(k_{\perp,c})$.
Furthermore, for the CEC, $\kappa\equiv\partial_{k_{\perp}^{2}}^{2}%
k_{\parallel,+}|_{k_{\perp,c}}$ gives the equivalent definition of Eq.
(\ref{DC}) for the differential curvature\cite{Carmo2016,GDOS2023}. To expand $k_{\mathbf{\parallel,}+}(k_{\perp})$ around
$k_{\mathbf{\perp,}c}$ up to quadratic order $k_{\mathbf{\parallel,}%
+}(k_{\perp})\approx k_{\parallel,c}+\kappa/2(k_{\perp}-k_{\perp,c})^{2}$,
this leads to%
\begin{equation}
\mathbf{g}(\varepsilon,\mathbf{r})\approx\frac{e^{ik_{\parallel,c}r}}%
{iv_{c}\sqrt{2\pi r}}\frac{e^{-i\pi/4}}{\sqrt{\left\vert\kappa\right\vert }%
}\left\vert u(\mathbf{k}_{c})\right\rangle \left\langle u(\mathbf{k}%
_{c})\right\vert,
\end{equation}
with $v_{c}=v_{\mathbf{\parallel}}(k_{\perp,c})$. Here, we have used the Gaussian
formula $\int_{-\infty}^{\infty}e^{-\alpha x^{2}}dx=\sqrt{\pi/\alpha}$ with
$\alpha\in\mathbb{C}$ and $\arg(\alpha)\in[-\pi/2,\pi/2]$. Using Eq. (\ref{DDOS}), the above equation can be rewritten as%
\begin{equation}
\mathbf{g}(\varepsilon,\mathbf{r})\approx-ie^{ik_{\parallel,c}r-i\pi/4}%
\sqrt{\frac{2\pi\rho}{v_{c}r}}\left\vert u(\mathbf{k}_{c})\right\rangle
\left\langle u(\mathbf{k}_{c})\right\vert , \label{GGF}%
\end{equation}
which presents the ultimate transparent expression with clear physics for
$\mathbf{g}(\varepsilon,\mathbf{r})$. Once the electronic structure is given,
it is convenient to construct $\mathbf{g}(\varepsilon,\mathbf{r})$, including
two processes; I) finding the stationary points $\mathbf{k}_{c}=(k_{x,c}%
,k_{y,c})$ on the CEC $E_{\lambda}(\mathbf{k})=\varepsilon$ satisfying
$\mathbf{v}_{c}\parallel\mathbf{r}$ with $\mathbf{v}_{c}$ the classic
velocity\textbf{, }this provides simultaneously $k_{\parallel,c}%
r=\mathbf{k}_{c}\cdot\mathbf{r}$ and $v_{c}=\left\vert \mathbf{v}%
_{c}\right\vert $\textbf{. }II) calculating $\rho(\varepsilon,\theta)$ through
Eq. (\ref{DDOS}). Eqs. (\ref{DDOS}) and (\ref{GGF}) are our main results, later
we use an example to show their specific applications.

\begin{figure}[ptbh]
\includegraphics[width=1.0\columnwidth,clip]{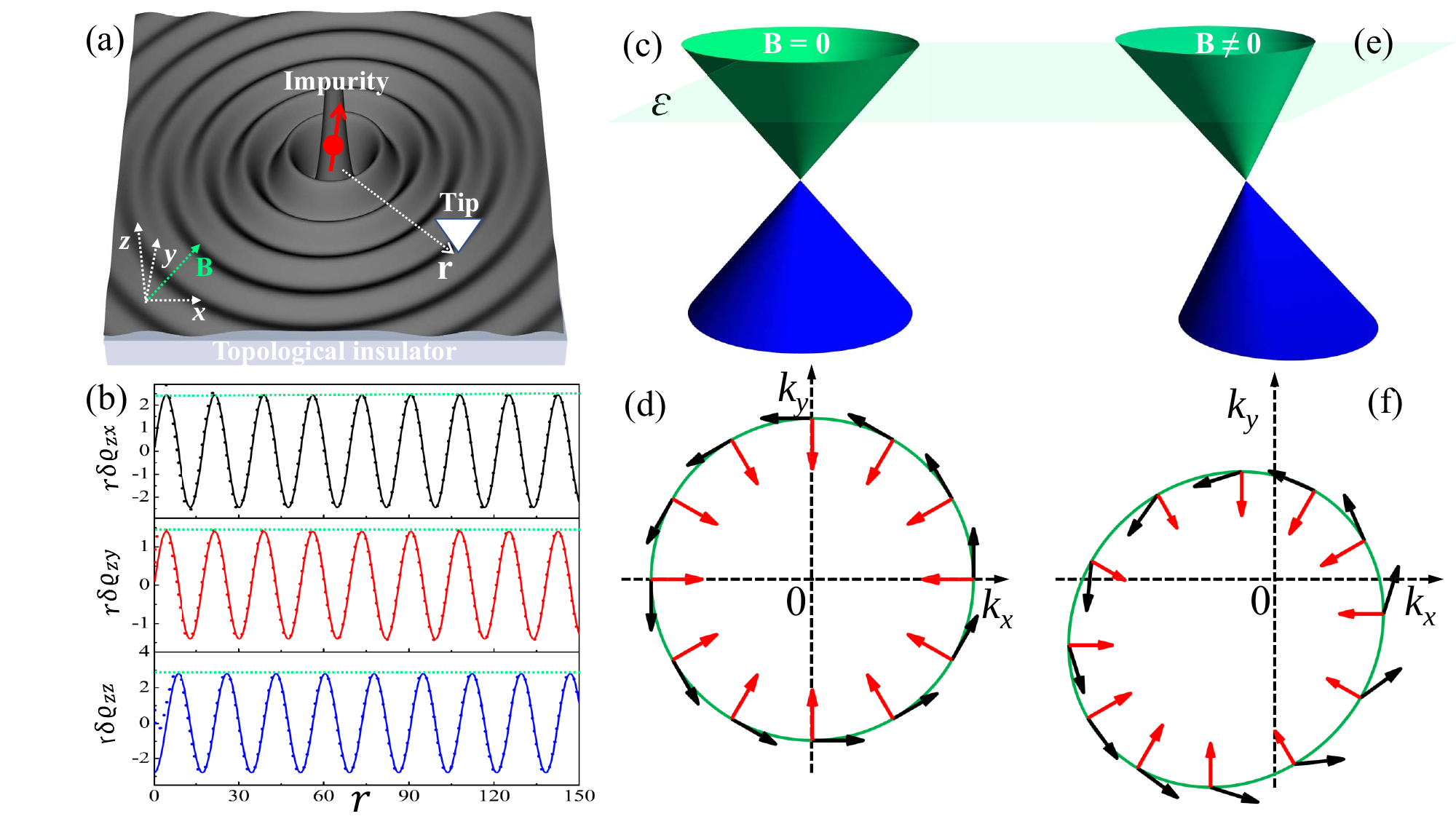}\caption{A protocol to
characterize topological surface states under an in-plane magnetic field. (a)
For the surface states (black) on the three-dimensional topological insulator
(light blue), a single magnetic impurity (red dot)\ at $\mathbf{r=0}$\ is
introduced intentionally. The impurity-induced Friedel oscillations (black
oscillation waves), namely the change of the local density of states (LDOS)
can be measured by spin-polarized scanning tunneling microscopy with the tip
at $\mathbf{r\neq0}$\ (white triangle), even under an in-plane magnetic field
$\mathbf{B}$. The inset shows the coordinate system with $\mathbf{B}$ lying in
the $x-y$ plane. (b) Along an arbitrary direction, the change of LDOS
$r\delta\varrho_{zx}$ (black lines),\ $r\delta\varrho_{zy}$ (red lines), and
$r\delta\varrho_{zz}$\ (blue lines) are shown, which do not decay by
accounting for the intrinsic $1/r$ decay. In each panel, solid line and dotted line
are, respectively, calculated by the $T$-matrix approach with and without Born
approximation, they agree very well when $r$ is longer than one oscillation
wavelength. The green lines in all panels help extract the amplitudes
$r\delta\varrho_{a,zx}$, $r\delta\varrho_{a,zy}$\ and $r\delta\varrho_{a,zz}$,
which determine the spin and curvature textures as shown in (d) and (f). In
(b), $v_{F}\equiv1$, $\theta_{\mathbf{r}}=\pi/6$, $\varepsilon=0.15$,
$\mathcal{V}=3$, and $t_{x}=t_{y}=0.3$. The in-plane $\mathbf{B}$ modifies the
electronic structure of topological surface states. The modification are
clearly by showing the electronic structures at $\mathbf{B}=0$ (c) and
$\mathbf{B}\neq0$ (e). Choosing the same Fermi levels $\varepsilon$ (light
green) in the (c) and (e), the constant-energy contour (CEC) is changed from
the circle (d) into ellipse (f). The CECs are fringed by the spin texture
(black)\ and the curvature texture (red), which can both be extracted through
the STM measurement (see the text). }%
\label{mtexture}%
\end{figure}

\textit{Characterizing the topological surface states at nonzero magnetic
field}.---The surface states of topological material (TISS)\ are the intuitive
reflection of the bulk-boundary correspondence\cite{RevModPhys.83.1057}, so it
is a basic requirement to characterize the electronic properties of TISS under
various fields. An in-plane magnetic field may lead to the drastic
modification of the electronic structure of TISS, which brings about the novel
effects such as the planar Hall
effect\cite{PhysRevB.101.041408,PhysRevB.103.155415} and the super-resonant
transport\cite{PhysRevLett.127.076601}. It is promising to corroborate the
modified TISS by an in-plane magnetic field, and there are two primary methods
to characterize the electronic structure, i.e., angle-resolved photoemission
spectroscopy (ARPES) and STM. But the conventional ARPES\ is incompatible with
magnetic field\cite{ZhengAIPX2018}, so STM measurements become obligatory and
there are a lot of STM experiments to visualize the surface states of
topological
materials\cite{nature08308,nature09189,ncomms6349,nmat4023,nphys2108,nphys3012,pnas.1424322112,PhysRevLett.103.266803,PhysRevLett.106.206805,PhysRevLett.112.136802,ZhengAIPX2018,s42254-021-00293-7}%
, but without considering the in-plane magnetic field. Here, we present a
protocol to measure the FO induced by the imperfection on the surface of
topological materials for characterizing the modified TISS [cf. Fig.
\ref{mtexture}(a)], and perform the theoretical analysis based on the ultimate
expression of the real-space GF. The modified TISS by an in-plane magnetic
field is described by the
Hamiltonian\cite{PhysRevB.101.041408,PhysRevLett.127.076601,GDOS2023}:%
\begin{equation}
\mathcal{H}(\mathbf{k})=v_{F}(k_{x}\sigma_{y}-k_{y}\sigma_{x})+\mathbf{t}%
\cdot\mathbf{k}.\label{Ham}%
\end{equation}
Here, $\sigma_{x,y}$ are Pauli matrices for the spin degrees of freedom,
$v_{F}$ is the velocity parameter and $\mathbf{t=(}t_{x},t_{y}\mathbf{)}$ is
the tilt vector induced by the applied in-plane magnetic filed $\mathbf{B}%
$\cite{PhysRevB.101.041408,PhysRevLett.127.076601}. The corresponding energy
dispersion and wavefunction are, respectively, $E_{\eta,\mathbf{k}}%
=t_{x}k_{x}+t_{y}k_{y}+\eta v_{F}k$, and $\left\vert u_{\eta}%
(\mathbf{k})\right\rangle =1/\sqrt{2}\left[
\begin{array}
[c]{cc}%
1 & \eta e^{i\Theta_{\mathbf{k}}}%
\end{array}
\right]  ^{\text{T}}$, where $\eta=\pm$ for the conductance and valence
bands, and $\Theta_{\mathbf{k}}=\eta\arg(-k_{y}+ik_{x})$ is the phase of wavefunction and also the azimuthal angle of the in-plane spin vector. The modification of TISS by an
in-plane magnetic field is clear by comparing the electronic structures at
$\mathbf{B}=\mathbf{0}$ [cf. Fig. \ref{mtexture}(c)]\ and $\mathbf{B}%
\neq\mathbf{0}$ [cf. Fig. \ref{mtexture}(e)]. Using the standard $T$-matrix
approach\cite{economou2006green,GDOS2023}, the FO or the change of local DOS (LDOS)
with energy and space resolution is given
as\cite{PhysRevB.81.233405,PhysRevB.85.125314}
\begin{equation}
\delta\varrho_{\alpha\beta}\left(  \varepsilon,\mathbf{r}\right)  =-\frac
{1}{\pi}\operatorname{Im}\mathrm{\operatorname*{Tr}}[\mathbf{g(}%
\varepsilon,\mathbf{r)T}_{\alpha}\mathbf{g(}\varepsilon,\mathbf{-r)}%
\sigma_{\beta}\mathbf{]},\label{LDOS}%
\end{equation}
where the $T$-matrix is expressed as%
\begin{equation}
\mathbf{T}_{\alpha}\left(  \varepsilon\right)  =\mathbf{V}_{\alpha}\left[
1-\mathbf{g}\left(  \varepsilon,\mathbf{0}\right)  \mathbf{V}_{\alpha}\right]
^{-1},\label{Tmatrix}%
\end{equation}
with $\mathbf{V}_{\alpha}$ from the impurity potential. Here, we use the
subscript $\alpha\in\{0,x,y,z\}$ with $\alpha=0$ and $\alpha\neq0$ for the
spin-unpolarized and spin-polarized imperfection or STM tip, respectively, so
$\delta\varrho_{\alpha\beta}\left(  \varepsilon,\mathbf{r}\right)  $ gives
$\beta$-resolved LDOS induced by $\alpha$-resolved imperfection. The
derivation of real-space GF is the prerequisite to analyse $\delta
\varrho_{\alpha\beta}\left(  \varepsilon,\mathbf{r}\right)  $, which is
usually hard for the model Hamiltonian\cite{PhysRevResearch.1.033188} and
time-consuming for the electronic structure of first-principle
calculations\cite{SmidstrupJPCM2019}. However, the ultimate expression of GF
(i.e., Eq. (\ref{GGF}))\ can be conveniently derived once the CEC is given,
even for the first-principle electronic structure. To utilize Eq. (\ref{GGF})
for the Hamiltonian Eq. (\ref{Ham}), we first need to derive the stationary
points $\mathbf{k}_{c}=(k_{x,c},k_{y,c})$. To consider the Fermi level in the
conduction band, the group velocity is $\mathbf{v}=(v_{x},v_{y})$ with
$v_{x,y}=t_{x,y}+v_{F}k_{x,y}/k$. On the stationary points,
$\mathbf{v\parallel r}$,\textbf{ }so $v_{y}/v_{x}=\tan\theta_{\mathbf{r}}$
which gives the equation for $\mathbf{k}$. Combining the energy dispersion
$E_{\eta,\mathbf{k}}=\varepsilon$ for given Fermi level $\varepsilon$, we
obtain%
\begin{subequations}
\begin{align}
k_{x,c}\left(  \theta_{\mathbf{r}}\right)   &  =\frac{\varepsilon}{v_{t}%
^{2}v_{m}}\left[  t_{x}t_{y}\sin\theta_{\mathbf{r}}+\left(  v_{F}^{2}%
-t_{y}^{2}\right)  \cos\theta_{\mathbf{r}}-v_{m}t_{x}\right]  \\
k_{y,c}\left(  \theta_{\mathbf{r}}\right)   &  =\frac{\varepsilon}{v_{t}%
^{2}v_{m}}\left[  t_{x}t_{y}\cos\theta_{\mathbf{r}}+(v_{F}^{2}-t_{x}^{2}%
)\sin\theta_{\mathbf{r}}-v_{m}t_{y}\right]  ,
\end{align}
where $v_{m}=\sqrt{v_{F}^{2}-t^{2}\sin^{2}(\phi-\theta_{\mathbf{r}})}\ $with
$\phi=\arg(t_{x}+it_{y})$, and $v_{t}^{2}=(v_{F}^{2}-t_{x}^{2}-t_{y}^{2})$.
Then one can obtain the classic velocity $\mathbf{v}_{c}(\mathbf{r})$ and then
its magnitude $v_{c}=\left\vert \mathbf{v}_{c}\right\vert =t\cos\left(
\theta_{\mathbf{r}}-\phi\right)  +v_{m}$. In addition, the curvature of CEC
corresponding to the energy dispersion is $\kappa\left(  \theta_{\mathbf{r}%
}\right)  =v_{m}^{3}/\varepsilon v_{F}^{2}$. According to Eq. (\ref{GGF}), the
explicit expression of GF is%
\end{subequations}
\begin{equation}
\mathbf{g}(\varepsilon,\pm\mathbf{r})=c_{\pm}e^{ik_{c,\pm}r}\left[
\begin{array}
[c]{cc}%
1 & e^{-i\Theta_{\mathbf{\pm}}}\\
e^{i\Theta_{\mathbf{\pm}}} & 1
\end{array}
\right]  ,
\end{equation}
where $c_{\pm}=-ie^{-i\pi/4}\sqrt{\pi\rho_{\pm}/(2v_{c,\pm}r)}$ with
$\rho_{\pm}=(4\pi^{2}v_{c,\pm}\kappa_{\pm})^{-1}$. Here, to account for
$\pm\mathbf{r}$\textbf{,} we define $k_{c,\pm}=k_{x,c}\left(  \theta
_{\pm\mathbf{r}}\right)  \cos\theta_{\pm\mathbf{r}}+k_{y,c}\left(  \theta
_{\pm\mathbf{r}}\right)  \sin\theta_{\pm\mathbf{r}}$, $\mathbf{v}_{c,\pm
}=\mathbf{v}_{c}(\pm\mathbf{r})$, $\kappa_{\pm}=\kappa\left(  \theta
_{\pm\mathbf{r}}\right)  $, and $\Theta_{c,\mathbf{\pm}}=\Theta
_{\mathbf{k}_{c,\pm}}$. To arrive at the explicit expression of GF,
the LDOS can be conveniently derived by incorporating the impurity potential\cite{GDOS2023}.
To intentionally introduce the single
imperfection\cite{437.full,s41586-019-1613-5,PhysRevLett.125.116804}, the
precise STM measurements will extract the information on the electronic
structure of the host system. In STM study, the conventional processes to
characterize the electronic structure are\cite{s42254-021-00293-7}: 1) probing
the LDOS with the resolution of the energy and position. 2) performing the
Fourier transform of LDOS to obtain experimental quasiparticle interference
(QPI). 3) Comparing the experimental QPI to theoretical QPI simulated based on
the first-principle or approximate electronic structures. To simulate the QPI,
$T$-matrix approach and joint DOS (JDOS) approximations are two available
methods, and the former is more accurate useful for the approximate electronic
structure while the latter is rather rough but used more widely due to its
support for the first-principle electronic structure. Our ultimate physical
expression of real-space GF favors the straightforward comparison of LDOS
between the experiments and simulations by using $T$-matrix approach even
based on the first-principles electronic structure, this should revolutionize
the theoretical understanding of STM measurement.

Back to Eq. (\ref{Tmatrix}), the furthermore calculations need the information
of impurity potential. In our focus, we intend to present the explicit
expressions for the LDOS $\delta\varrho_{\alpha\beta}\left(  \varepsilon
,\mathbf{r}\right)  $. To describe the imperfection as a $\delta$-function
potential $\mathbf{V}_{\alpha}\delta(\mathbf{r})$ with $\mathbf{V}_{\alpha
}=\mathcal{V}\sigma_{\alpha}$, $\delta\varrho_{\alpha\beta}\left(
\varepsilon,\mathbf{r}\right)  $ will have a explicit form in the Born
approximation\cite{economou2006green}. For example, adopting $\mathbf{V}%
_{\alpha}=\mathcal{V}\sigma_{z}$, we obtain
\begin{subequations}
\label{zLDOS}%
\begin{align}
\delta\varrho_{z0}  &  \approx\delta\varrho_{a,z0}\sin\left(  k_{+}r\right)
,\delta\varrho_{a,z0}\equiv\mathcal{C}\sin\left(  \Theta_{\mathbf{-}}%
-\Theta_{\mathbf{+}}\right)  ,\\
\delta\varrho_{zx}  &  \approx\delta\varrho_{a,zx}\sin\left(  k_{+}r\right)
,\delta\varrho_{a,zx}\equiv\mathcal{C}\left(  \sin\Theta_{\mathbf{-}}%
-\sin\Theta_{\mathbf{+}}\right)  ,\\
\delta\varrho_{zy}  &  \approx\delta\varrho_{a,zy}\sin\left(  k_{+}r\right)
,\delta\varrho_{a,zy}\equiv\mathcal{C}\left(  \cos\Theta_{\mathbf{+}}%
-\cos\Theta_{\mathbf{-}}\right)  ,\\
\delta\varrho_{zz}  &  \approx\delta\varrho_{a,zz}\cos\left(  k_{+}r\right)
,\delta\varrho_{a,zz}\equiv\mathcal{C}-\mathcal{C}\cos\left(  \Theta
_{\mathbf{-}}-\Theta_{\mathbf{+}}\right)  ,
\end{align}
where $\mathcal{C}=2\mathcal{V}c_{+}c_{-}/\pi=-(\mathcal{V}/r)\sqrt{\rho_{+}%
\rho_{-}/(v_{c,+}v_{c,-})}$, $k_{+}=k_{c,+}+k_{c,-}$ and $\delta
\varrho_{a,z\beta}$ is the amplitude of $\delta\varrho_{z\beta}$. If one
finishes the measurement of the LDOS induced by the designed imperfection, it
is convenient to compare with our theoretical simulations, which help
determine the Hamiltonian parameters of $v_{F}$ and $\mathbf{t}$\textbf{,} and
then all physical quantities through the Hamiltonian, e,g., CEC and the group
velocities $v_{c,\pm}$. This real-space comparison is usually impossible since
it is difficult to obtain the real-space GF. In particular, the impurity
potential strength $\mathcal{V}$ may be firstly extracted by performing the
same STM probe at zero magnetic field. In contrast to the QPI understanding of
STM measurement resorting to the reciprocal space, our theoretical
understanding is in real space and more intuitive.

The spin texture embodying in the phase of wavefunction is one of the most remarkable property of TISS, which usually
is probed by the ARPES\cite{nature08234,science.1167733,s42254-021-00293-7}.
To our knowledge, there is no experimental measurement of spin texture by
using STM. However, the GDOS, through the amplitude of GF, can realize the STM
measurement of the spin texture [cf. black arrows in Fig. \ref{mtexture}(d, f)] and even the
curvature texture [cf. red arrows in Fig. \ref{mtexture}(d, f)]. Thus, our explicit derivations
favors the extraction of the spin texture on the arbitrary position of CEC [cf. Fig. \ref{mtexture}(b)],
which surpasses previous theoretical simulations of STM
measurement\cite{s42254-021-00293-7}. $\delta\varrho_{a,z\beta}$ implies
\end{subequations}
\begin{subequations}
\label{texture}%
\begin{align}
\mathcal{C}  &  =\frac{\delta\varrho_{a,zx}^{2}+\delta\varrho
_{a,zy}^{2}}{2\delta\varrho_{a,zz}},\\
\cos(\Theta_{\mathbf{+}}-\Theta_{\mathbf{-}})  &  =1-\frac{\delta
\varrho_{a,zx}^{2}+\delta\varrho_{a,zy}^{2}}{2\mathcal{C}^{2}},\\
\cos(\Theta_{\mathbf{+}}+\Theta_{\mathbf{-}})  &  =\frac{\delta\varrho
_{a,zx}^{2}-\delta\varrho_{a,zy}^{2}}{\delta\varrho_{a,zx}^{2}+\delta
\varrho_{a,zy}^{2}}.
\end{align}
Therefore, the azimuthal angle $\Theta_{\mathbf{\pm}}$ of the spin states on
the stationary points $\mathbf{k}_{c,\pm}$\ can be solved. To shift the STM
tip around the imperfection, one can determine the spin texture on the
arbitrary points of the CEC. Nontrivially, the construction of spin texture
does not require the input of the information from\ the Hamiltonian except
assuming a spin-$1/2$ model.

Our theoretical derivations can go further. In particular, $\kappa_{+}%
=\kappa_{-}\equiv\kappa_{0}$ for Eq. (\ref{Ham}), so Eq. (\ref{texture}a) can
also be used to determine the curvature since \ \
\end{subequations}
\begin{equation}
\mathcal{C}=-\frac{\mathcal{V}}{4\pi^{2}r}\frac{1}{\kappa_{0}v_{c,+}v_{c,-}},
\end{equation}
and $v_{c,\pm}$ should be given through the CEC construction as discussed
previously. To fringe the curvature $\kappa_{0}$ on the CEC, the curvature
texture is given. In our opinion, it is a promising direction to explore the
geometrical curvature texture of the CEC, since it can be constructed through
the STM measurement.

\textit{Conclusions}.---In this study, we introduce a
geometrical amplitude for locally describing the electronic band structure,
namely the GDOS. Comparing to the conventional DOS $\rho(\varepsilon)$, the
GDOS $\rho(\varepsilon,\theta)$ obviously provides the information with the
higher complexity, as an elegant concept to understand the electronic
structure from the local perspective. The GDOS simplifies the construction of real-space GF as the basis for local responses and makes it attain the
ultimate expression with clear physics. In particular, in light of the
Fermi-surface property, the GDOS could be routinely calculated for the
first-principle band structures and furthermore favor the modular calculations
of ultimate expression of the real-space GF, this provide a straightforward
way to perform the theoretical simulations of STM experiment. In particular,
incorporating the ultimate real-space GF into the $T$-matrix approach helps
build the theoretical formulas for measuring the phase information (i.e., spin texture and curvature
texture of CEC) through the amplitudes of STM measurement. Finally, the GDOS should be an universal concept in periodic
system, such as photonic crystal, phononic crystal, and so on.  

\textit{Acknowledgements}.---This work was supported by the National Key
R$\&$D Program of China (Grant No. 2017YFA0303400), the NSFC (Grants No.
12174019, and No. 11774021), and the NSFC program for \textquotedblleft
Scientific Research Center\textquotedblright\ (Grant No. U1530401). S.H.Z. is
also supported by "Young Talent Program" of BUCT. We acknowledge the
computational support from the Beijing Computational Science Research Center (CSRC).


\begin{thebibliography}{54}
\expandafter\ifx\csname natexlab\endcsname\relax\def\natexlab#1{#1}\fi
\expandafter\ifx\csname bibnamefont\endcsname\relax
  \def\bibnamefont#1{#1}\fi
\expandafter\ifx\csname bibfnamefont\endcsname\relax
  \def\bibfnamefont#1{#1}\fi
\expandafter\ifx\csname citenamefont\endcsname\relax
  \def\citenamefont#1{#1}\fi
\expandafter\ifx\csname url\endcsname\relax
  \def\url#1{\texttt{#1}}\fi
\expandafter\ifx\csname urlprefix\endcsname\relax\def\urlprefix{URL }\fi
\providecommand{\bibinfo}[2]{#2}
\providecommand{\eprint}[2][]{\url{#2}}

\bibitem[{\citenamefont{Kittel}(2005)}]{Kittel2005}
\bibinfo{author}{\bibfnamefont{C.}~\bibnamefont{Kittel}},
  \emph{\bibinfo{title}{Indtroduction to Solid State Physics, 8th ed}}
  (\bibinfo{publisher}{Wiley, New York}, \bibinfo{year}{2005}).

\bibitem[{\citenamefont{Berry}(1984)}]{rspa.1984.0023}
\bibinfo{author}{\bibfnamefont{M.~V.} \bibnamefont{Berry}},
  \bibinfo{journal}{Proceedings of the Royal Society of London. A. Mathematical
  and Physical Sciences} \textbf{\bibinfo{volume}{392}}, \bibinfo{pages}{45}
  (\bibinfo{year}{1984}).

\bibitem[{\citenamefont{{Cohen, Eliahu and Larocque, Hugo and Bouchard,
  Frederic and Nejadsattari, Farshad and Gefen, Yuval and Karimi,
  Ebrahim}}(2019)}]{s42254-019-0071-1}
\bibinfo{author}{\bibnamefont{{Cohen, Eliahu and Larocque, Hugo and Bouchard,
  Frederic and Nejadsattari, Farshad and Gefen, Yuval and Karimi, Ebrahim}}},
  \bibinfo{journal}{Nature Reviews Physics} \textbf{\bibinfo{volume}{1}},
  \bibinfo{pages}{437} (\bibinfo{year}{2019}).

\bibitem[{\citenamefont{Xiao et~al.}(2010)\citenamefont{Xiao, Chang, and
  Niu}}]{RevModPhys.82.1959}
\bibinfo{author}{\bibfnamefont{D.}~\bibnamefont{Xiao}},
  \bibinfo{author}{\bibfnamefont{M.-C.} \bibnamefont{Chang}}, \bibnamefont{and}
  \bibinfo{author}{\bibfnamefont{Q.}~\bibnamefont{Niu}}, \bibinfo{journal}{Rev.
  Mod. Phys.} \textbf{\bibinfo{volume}{82}}, \bibinfo{pages}{1959}
  (\bibinfo{year}{2010}).

\bibitem[{\citenamefont{Vanderbilt}(2018)}]{vanderbilt2018}
\bibinfo{author}{\bibfnamefont{D.}~\bibnamefont{Vanderbilt}},
  \emph{\bibinfo{title}{Berry Phases in Electronic Structure Theory: Electric
  Polarization, Orbital Magnetization and Topological Insulators}}
  (\bibinfo{publisher}{Cambridge University Press}, \bibinfo{year}{2018}).

\bibitem[{\citenamefont{Yin et~al.}(2021)\citenamefont{Yin, Pan, and
  Zahid~Hasan}}]{s42254-021-00293-7}
\bibinfo{author}{\bibfnamefont{J.-X.} \bibnamefont{Yin}},
  \bibinfo{author}{\bibfnamefont{S.~H.} \bibnamefont{Pan}}, \bibnamefont{and}
  \bibinfo{author}{\bibfnamefont{M.}~\bibnamefont{Zahid~Hasan}},
  \bibinfo{journal}{Nature Reviews Physics} \textbf{\bibinfo{volume}{3}},
  \bibinfo{pages}{249} (\bibinfo{year}{2021}).

\bibitem[{\citenamefont{Zhang et~al.}(2022)\citenamefont{Zhang, Yi, and
  Xu}}]{ZhangJPCM2022}
\bibinfo{author}{\bibfnamefont{C.}~\bibnamefont{Zhang}},
  \bibinfo{author}{\bibfnamefont{Z.}~\bibnamefont{Yi}}, \bibnamefont{and}
  \bibinfo{author}{\bibfnamefont{W.}~\bibnamefont{Xu}},
  \bibinfo{journal}{Materials Futures} \textbf{\bibinfo{volume}{1}},
  \bibinfo{pages}{032301} (\bibinfo{year}{2022}).

\bibitem[{\citenamefont{Dutreix et~al.}(2019)\citenamefont{Dutreix,
  Gonzalez-Herrero, Brihuega, Katsnelson, Chapelier, and
  Renard}}]{s41586-019-1613-5}
\bibinfo{author}{\bibfnamefont{C.}~\bibnamefont{Dutreix}},
  \bibinfo{author}{\bibfnamefont{H.}~\bibnamefont{Gonzalez-Herrero}},
  \bibinfo{author}{\bibfnamefont{I.}~\bibnamefont{Brihuega}},
  \bibinfo{author}{\bibfnamefont{M.~I.} \bibnamefont{Katsnelson}},
  \bibinfo{author}{\bibfnamefont{C.}~\bibnamefont{Chapelier}},
  \bibnamefont{and} \bibinfo{author}{\bibfnamefont{V.~T.}
  \bibnamefont{Renard}}, \bibinfo{journal}{Nature}
  \textbf{\bibinfo{volume}{574}}, \bibinfo{pages}{219} (\bibinfo{year}{2019}).

\bibitem[{\citenamefont{Zhang et~al.}(2020)\citenamefont{Zhang, Su, and
  He}}]{PhysRevLett.125.116804}
\bibinfo{author}{\bibfnamefont{Y.}~\bibnamefont{Zhang}},
  \bibinfo{author}{\bibfnamefont{Y.}~\bibnamefont{Su}}, \bibnamefont{and}
  \bibinfo{author}{\bibfnamefont{L.}~\bibnamefont{He}}, \bibinfo{journal}{Phys.
  Rev. Lett.} \textbf{\bibinfo{volume}{125}}, \bibinfo{pages}{116804}
  (\bibinfo{year}{2020}).

\bibitem[{\citenamefont{Dutreix et~al.}(2021)\citenamefont{Dutreix,
  Gonz\'alez-Herrero, Brihuega, Katsnelson, Chapelier, and
  Renard}}]{CRPHYS2021}
\bibinfo{author}{\bibfnamefont{C.}~\bibnamefont{Dutreix}},
  \bibinfo{author}{\bibfnamefont{H.}~\bibnamefont{Gonz\'alez-Herrero}},
  \bibinfo{author}{\bibfnamefont{I.}~\bibnamefont{Brihuega}},
  \bibinfo{author}{\bibfnamefont{M.~I.} \bibnamefont{Katsnelson}},
  \bibinfo{author}{\bibfnamefont{C.}~\bibnamefont{Chapelier}},
  \bibnamefont{and} \bibinfo{author}{\bibfnamefont{V.~T.}
  \bibnamefont{Renard}}, \bibinfo{journal}{Comptes Rendus. Physique}
  \textbf{\bibinfo{volume}{22}}, \bibinfo{pages}{133} (\bibinfo{year}{2021}).

\bibitem[{\citenamefont{Castro~Neto et~al.}(2009)\citenamefont{Castro~Neto,
  Guinea, Peres, Novoselov, and Geim}}]{RevModPhys.81.109}
\bibinfo{author}{\bibfnamefont{A.~H.} \bibnamefont{Castro~Neto}},
  \bibinfo{author}{\bibfnamefont{F.}~\bibnamefont{Guinea}},
  \bibinfo{author}{\bibfnamefont{N.~M.~R.} \bibnamefont{Peres}},
  \bibinfo{author}{\bibfnamefont{K.~S.} \bibnamefont{Novoselov}},
  \bibnamefont{and} \bibinfo{author}{\bibfnamefont{A.~K.} \bibnamefont{Geim}},
  \bibinfo{journal}{Rev. Mod. Phys.} \textbf{\bibinfo{volume}{81}},
  \bibinfo{pages}{109} (\bibinfo{year}{2009}).

\bibitem[{\citenamefont{Han et~al.}(2014)\citenamefont{Han, Kawakami, Gmitra,
  and Fabian}}]{HanNatureNanotechnology2014}
\bibinfo{author}{\bibfnamefont{W.}~\bibnamefont{Han}},
  \bibinfo{author}{\bibfnamefont{R.~K.} \bibnamefont{Kawakami}},
  \bibinfo{author}{\bibfnamefont{M.}~\bibnamefont{Gmitra}}, \bibnamefont{and}
  \bibinfo{author}{\bibfnamefont{J.}~\bibnamefont{Fabian}},
  \bibinfo{journal}{Nature Nanotechnology} \textbf{\bibinfo{volume}{9}},
  \bibinfo{pages}{794} (\bibinfo{year}{2014}).

\bibitem[{\citenamefont{Wang et~al.}(2018)\citenamefont{Wang, Ren, Yan, Jiang,
  Sha, and Shan}}]{s11467-018-0859-y}
\bibinfo{author}{\bibfnamefont{R.}~\bibnamefont{Wang}},
  \bibinfo{author}{\bibfnamefont{X.-G.} \bibnamefont{Ren}},
  \bibinfo{author}{\bibfnamefont{Z.}~\bibnamefont{Yan}},
  \bibinfo{author}{\bibfnamefont{L.-J.} \bibnamefont{Jiang}},
  \bibinfo{author}{\bibfnamefont{W.~E.~I.} \bibnamefont{Sha}},
  \bibnamefont{and} \bibinfo{author}{\bibfnamefont{G.-C.} \bibnamefont{Shan}},
  \bibinfo{journal}{Frontiers of Physics} \textbf{\bibinfo{volume}{14}},
  \bibinfo{pages}{13603} (\bibinfo{year}{2018}).

\bibitem[{\citenamefont{Liu et~al.}(2016)\citenamefont{Liu, Weiss, Duan, Cheng,
  Huang, and Duan}}]{natrevmats.2016.42}
\bibinfo{author}{\bibfnamefont{Y.}~\bibnamefont{Liu}},
  \bibinfo{author}{\bibfnamefont{N.~O.} \bibnamefont{Weiss}},
  \bibinfo{author}{\bibfnamefont{X.}~\bibnamefont{Duan}},
  \bibinfo{author}{\bibfnamefont{H.-C.} \bibnamefont{Cheng}},
  \bibinfo{author}{\bibfnamefont{Y.}~\bibnamefont{Huang}}, \bibnamefont{and}
  \bibinfo{author}{\bibfnamefont{X.}~\bibnamefont{Duan}},
  \bibinfo{journal}{Nature Reviews Materials} \textbf{\bibinfo{volume}{1}},
  \bibinfo{pages}{16042} (\bibinfo{year}{2016}).

\bibitem[{\citenamefont{Akinwande et~al.}(2019)\citenamefont{Akinwande,
  Huyghebaert, Wang, Serna, Goossens, Li, Wong, and
  Koppens}}]{s41586-019-1573-9}
\bibinfo{author}{\bibfnamefont{D.}~\bibnamefont{Akinwande}},
  \bibinfo{author}{\bibfnamefont{C.}~\bibnamefont{Huyghebaert}},
  \bibinfo{author}{\bibfnamefont{C.-H.} \bibnamefont{Wang}},
  \bibinfo{author}{\bibfnamefont{M.~I.} \bibnamefont{Serna}},
  \bibinfo{author}{\bibfnamefont{S.}~\bibnamefont{Goossens}},
  \bibinfo{author}{\bibfnamefont{L.-J.} \bibnamefont{Li}},
  \bibinfo{author}{\bibfnamefont{H.-S.~P.} \bibnamefont{Wong}},
  \bibnamefont{and} \bibinfo{author}{\bibfnamefont{F.~H.~L.}
  \bibnamefont{Koppens}}, \bibinfo{journal}{Nature}
  \textbf{\bibinfo{volume}{573}}, \bibinfo{pages}{507} (\bibinfo{year}{2019}).

\bibitem[{\citenamefont{Novoselov et~al.}(2016)\citenamefont{Novoselov,
  Mishchenko, Carvalho, and Castro~Neto}}]{aac9439}
\bibinfo{author}{\bibfnamefont{K.~S.} \bibnamefont{Novoselov}},
  \bibinfo{author}{\bibfnamefont{A.}~\bibnamefont{Mishchenko}},
  \bibinfo{author}{\bibfnamefont{A.}~\bibnamefont{Carvalho}}, \bibnamefont{and}
  \bibinfo{author}{\bibfnamefont{A.~H.} \bibnamefont{Castro~Neto}},
  \bibinfo{journal}{Science} \textbf{\bibinfo{volume}{353}},
  \bibinfo{pages}{aac9439} (\bibinfo{year}{2016}).

\bibitem[{\citenamefont{Avsar et~al.}(2020)\citenamefont{Avsar, Ochoa, Guinea,
  \"Ozyilmaz, van Wees, and Vera-Marun}}]{RevModPhys.92.021003}
\bibinfo{author}{\bibfnamefont{A.}~\bibnamefont{Avsar}},
  \bibinfo{author}{\bibfnamefont{H.}~\bibnamefont{Ochoa}},
  \bibinfo{author}{\bibfnamefont{F.}~\bibnamefont{Guinea}},
  \bibinfo{author}{\bibfnamefont{B.}~\bibnamefont{\"Ozyilmaz}},
  \bibinfo{author}{\bibfnamefont{B.~J.} \bibnamefont{van Wees}},
  \bibnamefont{and} \bibinfo{author}{\bibfnamefont{I.~J.}
  \bibnamefont{Vera-Marun}}, \bibinfo{journal}{Rev. Mod. Phys.}
  \textbf{\bibinfo{volume}{92}}, \bibinfo{pages}{021003}
  (\bibinfo{year}{2020}).

\bibitem[{\citenamefont{Toriyama et~al.}(2022)\citenamefont{Toriyama, Ganose,
  Dylla, Anand, Park, Brod, Munro, Persson, Jain, and
  Snyder}}]{1-s2.0-S277294942200002X-main}
\bibinfo{author}{\bibfnamefont{M.~Y.} \bibnamefont{Toriyama}},
  \bibinfo{author}{\bibfnamefont{A.~M.} \bibnamefont{Ganose}},
  \bibinfo{author}{\bibfnamefont{M.}~\bibnamefont{Dylla}},
  \bibinfo{author}{\bibfnamefont{S.}~\bibnamefont{Anand}},
  \bibinfo{author}{\bibfnamefont{J.}~\bibnamefont{Park}},
  \bibinfo{author}{\bibfnamefont{M.~K.} \bibnamefont{Brod}},
  \bibinfo{author}{\bibfnamefont{J.~M.} \bibnamefont{Munro}},
  \bibinfo{author}{\bibfnamefont{K.~A.} \bibnamefont{Persson}},
  \bibinfo{author}{\bibfnamefont{A.}~\bibnamefont{Jain}}, \bibnamefont{and}
  \bibinfo{author}{\bibfnamefont{G.~J.} \bibnamefont{Snyder}},
  \bibinfo{journal}{Materials Today Electronics} \textbf{\bibinfo{volume}{1}},
  \bibinfo{pages}{100002} (\bibinfo{year}{2022}).

\bibitem[{\citenamefont{Simon et~al.}(2011)\citenamefont{Simon, Bena, Vonau,
  Cranney, and Aubel}}]{SimonJPDAP2011}
\bibinfo{author}{\bibfnamefont{L.}~\bibnamefont{Simon}},
  \bibinfo{author}{\bibfnamefont{C.}~\bibnamefont{Bena}},
  \bibinfo{author}{\bibfnamefont{F.}~\bibnamefont{Vonau}},
  \bibinfo{author}{\bibfnamefont{M.}~\bibnamefont{Cranney}}, \bibnamefont{and}
  \bibinfo{author}{\bibfnamefont{D.}~\bibnamefont{Aubel}}, \bibinfo{journal}{J.
  Phys. D: Appl. Phys.} \textbf{\bibinfo{volume}{44}}, \bibinfo{pages}{464010}
  (\bibinfo{year}{2011}).

\bibitem[{\citenamefont{Bena}(2016)}]{Bena2016}
\bibinfo{author}{\bibfnamefont{C.}~\bibnamefont{Bena}},
  \bibinfo{journal}{Comptes Rendus Physique} \textbf{\bibinfo{volume}{17}},
  \bibinfo{pages}{302} (\bibinfo{year}{2016}).

\bibitem[{\citenamefont{Chen et~al.}(2017)\citenamefont{Chen, Cheng, and
  Wu}}]{ChenJPCM2017}
\bibinfo{author}{\bibfnamefont{L.}~\bibnamefont{Chen}},
  \bibinfo{author}{\bibfnamefont{P.}~\bibnamefont{Cheng}}, \bibnamefont{and}
  \bibinfo{author}{\bibfnamefont{K.}~\bibnamefont{Wu}}, \bibinfo{journal}{J.
  Phys. Condens. Matter} \textbf{\bibinfo{volume}{29}}, \bibinfo{pages}{103001}
  (\bibinfo{year}{2017}).

\bibitem[{\citenamefont{Avraham et~al.}(2018)\citenamefont{Avraham, Reiner,
  Kumar-Nayak, Morali, Batabyal, Yan, and Beidenkopf}}]{adma.201707628}
\bibinfo{author}{\bibfnamefont{N.}~\bibnamefont{Avraham}},
  \bibinfo{author}{\bibfnamefont{J.}~\bibnamefont{Reiner}},
  \bibinfo{author}{\bibfnamefont{A.}~\bibnamefont{Kumar-Nayak}},
  \bibinfo{author}{\bibfnamefont{N.}~\bibnamefont{Morali}},
  \bibinfo{author}{\bibfnamefont{R.}~\bibnamefont{Batabyal}},
  \bibinfo{author}{\bibfnamefont{B.}~\bibnamefont{Yan}}, \bibnamefont{and}
  \bibinfo{author}{\bibfnamefont{H.}~\bibnamefont{Beidenkopf}},
  \bibinfo{journal}{Advanced Materials} \textbf{\bibinfo{volume}{30}},
  \bibinfo{pages}{1707628} (\bibinfo{year}{2018}).

\bibitem[{\citenamefont{Zheng and Hasan}(2018)}]{ZhengAIPX2018}
\bibinfo{author}{\bibfnamefont{H.}~\bibnamefont{Zheng}} \bibnamefont{and}
  \bibinfo{author}{\bibfnamefont{M.~Z.} \bibnamefont{Hasan}},
  \bibinfo{journal}{Advances in Physics: X} \textbf{\bibinfo{volume}{3}},
  \bibinfo{pages}{1466661} (\bibinfo{year}{2018}).

\bibitem[{\citenamefont{Lin et~al.}(2020)\citenamefont{Lin, Kawakami, Arafune,
  Minamitani, and Takagi}}]{LinJPCM2020}
\bibinfo{author}{\bibfnamefont{C.-L.} \bibnamefont{Lin}},
  \bibinfo{author}{\bibfnamefont{N.}~\bibnamefont{Kawakami}},
  \bibinfo{author}{\bibfnamefont{R.}~\bibnamefont{Arafune}},
  \bibinfo{author}{\bibfnamefont{E.}~\bibnamefont{Minamitani}},
  \bibnamefont{and} \bibinfo{author}{\bibfnamefont{N.}~\bibnamefont{Takagi}},
  \bibinfo{journal}{Journal of Physics: Condensed Matter}
  \textbf{\bibinfo{volume}{32}}, \bibinfo{pages}{243001}
  (\bibinfo{year}{2020}).

\bibitem[{\citenamefont{Sato et~al.}(2010)\citenamefont{Sato, Bergqvist,
  Kudrnovsk\'y, Dederichs, Eriksson, Turek, Sanyal, Bouzerar, Katayama-Yoshida,
  Dinh et~al.}}]{RevModPhys.82.1633}
\bibinfo{author}{\bibfnamefont{K.}~\bibnamefont{Sato}},
  \bibinfo{author}{\bibfnamefont{L.}~\bibnamefont{Bergqvist}},
  \bibinfo{author}{\bibfnamefont{J.}~\bibnamefont{Kudrnovsk\'y}},
  \bibinfo{author}{\bibfnamefont{P.~H.} \bibnamefont{Dederichs}},
  \bibinfo{author}{\bibfnamefont{O.}~\bibnamefont{Eriksson}},
  \bibinfo{author}{\bibfnamefont{I.}~\bibnamefont{Turek}},
  \bibinfo{author}{\bibfnamefont{B.}~\bibnamefont{Sanyal}},
  \bibinfo{author}{\bibfnamefont{G.}~\bibnamefont{Bouzerar}},
  \bibinfo{author}{\bibfnamefont{H.}~\bibnamefont{Katayama-Yoshida}},
  \bibinfo{author}{\bibfnamefont{V.~A.} \bibnamefont{Dinh}},
  \bibnamefont{et~al.}, \bibinfo{journal}{Rev. Mod. Phys.}
  \textbf{\bibinfo{volume}{82}}, \bibinfo{pages}{1633} (\bibinfo{year}{2010}).

\bibitem[{\citenamefont{Power and Ferreira}(2013)}]{cryst3010049}
\bibinfo{author}{\bibfnamefont{S.~R.} \bibnamefont{Power}} \bibnamefont{and}
  \bibinfo{author}{\bibfnamefont{M.~S.} \bibnamefont{Ferreira}},
  \bibinfo{journal}{Crystals} \textbf{\bibinfo{volume}{3}}, \bibinfo{pages}{49}
  (\bibinfo{year}{2013}).

\bibitem[{\citenamefont{Settnes et~al.}(2014)\citenamefont{Settnes, Power,
  Petersen, and Jauho}}]{SettnesPRL2014}
\bibinfo{author}{\bibfnamefont{M.}~\bibnamefont{Settnes}},
  \bibinfo{author}{\bibfnamefont{S.~R.} \bibnamefont{Power}},
  \bibinfo{author}{\bibfnamefont{D.~H.} \bibnamefont{Petersen}},
  \bibnamefont{and} \bibinfo{author}{\bibfnamefont{A.-P.} \bibnamefont{Jauho}},
  \bibinfo{journal}{Phys. Rev. Lett.} \textbf{\bibinfo{volume}{112}},
  \bibinfo{pages}{096801} (\bibinfo{year}{2014}).

\bibitem[{\citenamefont{Zheng et~al.}(2020)\citenamefont{Zheng, Duan, Wang, Li,
  Deng, and Wang}}]{PhysRevB.101.041408}
\bibinfo{author}{\bibfnamefont{S.-H.} \bibnamefont{Zheng}},
  \bibinfo{author}{\bibfnamefont{H.-J.} \bibnamefont{Duan}},
  \bibinfo{author}{\bibfnamefont{J.-K.} \bibnamefont{Wang}},
  \bibinfo{author}{\bibfnamefont{J.-Y.} \bibnamefont{Li}},
  \bibinfo{author}{\bibfnamefont{M.-X.} \bibnamefont{Deng}}, \bibnamefont{and}
  \bibinfo{author}{\bibfnamefont{R.-Q.} \bibnamefont{Wang}},
  \bibinfo{journal}{Phys. Rev. B} \textbf{\bibinfo{volume}{101}},
  \bibinfo{pages}{041408} (\bibinfo{year}{2020}).

\bibitem[{\citenamefont{Rao et~al.}(2021)\citenamefont{Rao, Zhou, Wu, Duan,
  Deng, and Wang}}]{PhysRevB.103.155415}
\bibinfo{author}{\bibfnamefont{W.}~\bibnamefont{Rao}},
  \bibinfo{author}{\bibfnamefont{Y.-L.} \bibnamefont{Zhou}},
  \bibinfo{author}{\bibfnamefont{Y.-j.} \bibnamefont{Wu}},
  \bibinfo{author}{\bibfnamefont{H.-J.} \bibnamefont{Duan}},
  \bibinfo{author}{\bibfnamefont{M.-X.} \bibnamefont{Deng}}, \bibnamefont{and}
  \bibinfo{author}{\bibfnamefont{R.-Q.} \bibnamefont{Wang}},
  \bibinfo{journal}{Phys. Rev. B} \textbf{\bibinfo{volume}{103}},
  \bibinfo{pages}{155415} (\bibinfo{year}{2021}).

\bibitem[{\citenamefont{Zhang et~al.}(2021)\citenamefont{Zhang, Li, Pe\~na
  Benitez, Sur\'owka, Moessner, Molenkamp, and
  Trauzettel}}]{PhysRevLett.127.076601}
\bibinfo{author}{\bibfnamefont{S.-B.} \bibnamefont{Zhang}},
  \bibinfo{author}{\bibfnamefont{C.-A.} \bibnamefont{Li}},
  \bibinfo{author}{\bibfnamefont{F.}~\bibnamefont{Pe\~na Benitez}},
  \bibinfo{author}{\bibfnamefont{P.}~\bibnamefont{Sur\'owka}},
  \bibinfo{author}{\bibfnamefont{R.}~\bibnamefont{Moessner}},
  \bibinfo{author}{\bibfnamefont{L.~W.} \bibnamefont{Molenkamp}},
  \bibnamefont{and}
  \bibinfo{author}{\bibfnamefont{B.}~\bibnamefont{Trauzettel}},
  \bibinfo{journal}{Phys. Rev. Lett.} \textbf{\bibinfo{volume}{127}},
  \bibinfo{pages}{076601} (\bibinfo{year}{2021}).

\bibitem[{\citenamefont{do~Carmo}(2016)}]{Carmo2016}
\bibinfo{author}{\bibfnamefont{M.~P.} \bibnamefont{do~Carmo}},
  \emph{\bibinfo{title}{Differential Geometry of Curves and Surfaces}}
  (\bibinfo{publisher}{Dover Publications, New York}, \bibinfo{year}{2016}).

\bibitem[{GDO()}]{GDOS2023}
\bibinfo{journal}{Supplementary Materials}  (????).

\bibitem[{\citenamefont{Economou}(1983)}]{Economou1983}
\bibinfo{author}{\bibfnamefont{E.~N.} \bibnamefont{Economou}},
  \emph{\bibinfo{title}{Green’s Functions in Quantum Physics}}
  (\bibinfo{publisher}{Springer, Berlin}, \bibinfo{year}{1983}).

\bibitem[{\citenamefont{Roth et~al.}(1966)\citenamefont{Roth, Zeiger, and
  Kaplan}}]{PhysRev.149.519}
\bibinfo{author}{\bibfnamefont{L.~M.} \bibnamefont{Roth}},
  \bibinfo{author}{\bibfnamefont{H.~J.} \bibnamefont{Zeiger}},
  \bibnamefont{and} \bibinfo{author}{\bibfnamefont{T.~A.}
  \bibnamefont{Kaplan}}, \bibinfo{journal}{Phys. Rev.}
  \textbf{\bibinfo{volume}{149}}, \bibinfo{pages}{519} (\bibinfo{year}{1966}).

\bibitem[{\citenamefont{Lounis et~al.}(2011)\citenamefont{Lounis, Zahn,
  Weismann, Wenderoth, Ulbrich, Mertig, Dederichs, and
  Bl\"ugel}}]{PhysRevB.83.035427}
\bibinfo{author}{\bibfnamefont{S.}~\bibnamefont{Lounis}},
  \bibinfo{author}{\bibfnamefont{P.}~\bibnamefont{Zahn}},
  \bibinfo{author}{\bibfnamefont{A.}~\bibnamefont{Weismann}},
  \bibinfo{author}{\bibfnamefont{M.}~\bibnamefont{Wenderoth}},
  \bibinfo{author}{\bibfnamefont{R.~G.} \bibnamefont{Ulbrich}},
  \bibinfo{author}{\bibfnamefont{I.}~\bibnamefont{Mertig}},
  \bibinfo{author}{\bibfnamefont{P.~H.} \bibnamefont{Dederichs}},
  \bibnamefont{and} \bibinfo{author}{\bibfnamefont{S.}~\bibnamefont{Bl\"ugel}},
  \bibinfo{journal}{Phys. Rev. B} \textbf{\bibinfo{volume}{83}},
  \bibinfo{pages}{035427} (\bibinfo{year}{2011}).

\bibitem[{\citenamefont{Liu et~al.}(2012)\citenamefont{Liu, Qi, and
  Zhang}}]{PhysRevB.85.125314}
\bibinfo{author}{\bibfnamefont{Q.}~\bibnamefont{Liu}},
  \bibinfo{author}{\bibfnamefont{X.-L.} \bibnamefont{Qi}}, \bibnamefont{and}
  \bibinfo{author}{\bibfnamefont{S.-C.} \bibnamefont{Zhang}},
  \bibinfo{journal}{Phys. Rev. B} \textbf{\bibinfo{volume}{85}},
  \bibinfo{pages}{125314} (\bibinfo{year}{2012}).

\bibitem[{\citenamefont{Qi and Zhang}(2011)}]{RevModPhys.83.1057}
\bibinfo{author}{\bibfnamefont{X.-L.} \bibnamefont{Qi}} \bibnamefont{and}
  \bibinfo{author}{\bibfnamefont{S.-C.} \bibnamefont{Zhang}},
  \bibinfo{journal}{Rev. Mod. Phys.} \textbf{\bibinfo{volume}{83}},
  \bibinfo{pages}{1057} (\bibinfo{year}{2011}).

\bibitem[{\citenamefont{Roushan et~al.}(2009)\citenamefont{Roushan, Seo,
  Parker, Hor, Hsieh, Qian, Richardella, Hasan, Cava, and
  Yazdani}}]{nature08308}
\bibinfo{author}{\bibfnamefont{P.}~\bibnamefont{Roushan}},
  \bibinfo{author}{\bibfnamefont{J.}~\bibnamefont{Seo}},
  \bibinfo{author}{\bibfnamefont{C.~V.} \bibnamefont{Parker}},
  \bibinfo{author}{\bibfnamefont{Y.~S.} \bibnamefont{Hor}},
  \bibinfo{author}{\bibfnamefont{D.}~\bibnamefont{Hsieh}},
  \bibinfo{author}{\bibfnamefont{D.}~\bibnamefont{Qian}},
  \bibinfo{author}{\bibfnamefont{A.}~\bibnamefont{Richardella}},
  \bibinfo{author}{\bibfnamefont{M.~Z.} \bibnamefont{Hasan}},
  \bibinfo{author}{\bibfnamefont{R.~J.} \bibnamefont{Cava}}, \bibnamefont{and}
  \bibinfo{author}{\bibfnamefont{A.}~\bibnamefont{Yazdani}},
  \bibinfo{journal}{Nature} \textbf{\bibinfo{volume}{460}},
  \bibinfo{pages}{1106} (\bibinfo{year}{2009}).

\bibitem[{\citenamefont{Seo et~al.}(2010)\citenamefont{Seo, Roushan,
  Beidenkopf, Hor, Cava, and Yazdani}}]{nature09189}
\bibinfo{author}{\bibfnamefont{J.}~\bibnamefont{Seo}},
  \bibinfo{author}{\bibfnamefont{P.}~\bibnamefont{Roushan}},
  \bibinfo{author}{\bibfnamefont{H.}~\bibnamefont{Beidenkopf}},
  \bibinfo{author}{\bibfnamefont{Y.~S.} \bibnamefont{Hor}},
  \bibinfo{author}{\bibfnamefont{R.~J.} \bibnamefont{Cava}}, \bibnamefont{and}
  \bibinfo{author}{\bibfnamefont{A.}~\bibnamefont{Yazdani}},
  \bibinfo{journal}{Nature} \textbf{\bibinfo{volume}{466}},
  \bibinfo{pages}{343} (\bibinfo{year}{2010}).

\bibitem[{\citenamefont{Sessi et~al.}(2014)\citenamefont{Sessi, Reis, Bathon,
  Kokh, Tereshchenko, and Bode}}]{ncomms6349}
\bibinfo{author}{\bibfnamefont{P.}~\bibnamefont{Sessi}},
  \bibinfo{author}{\bibfnamefont{F.}~\bibnamefont{Reis}},
  \bibinfo{author}{\bibfnamefont{T.}~\bibnamefont{Bathon}},
  \bibinfo{author}{\bibfnamefont{K.~A.} \bibnamefont{Kokh}},
  \bibinfo{author}{\bibfnamefont{O.~E.} \bibnamefont{Tereshchenko}},
  \bibnamefont{and} \bibinfo{author}{\bibfnamefont{M.}~\bibnamefont{Bode}},
  \bibinfo{journal}{Nature Communications} \textbf{\bibinfo{volume}{5}},
  \bibinfo{pages}{5349} (\bibinfo{year}{2014}).

\bibitem[{\citenamefont{Jeon et~al.}(2014)\citenamefont{Jeon, Zhou, Gyenis,
  Feldman, Kimchi, Potter, Gibson, Cava, Vishwanath, and Yazdani}}]{nmat4023}
\bibinfo{author}{\bibfnamefont{S.}~\bibnamefont{Jeon}},
  \bibinfo{author}{\bibfnamefont{B.~B.} \bibnamefont{Zhou}},
  \bibinfo{author}{\bibfnamefont{A.}~\bibnamefont{Gyenis}},
  \bibinfo{author}{\bibfnamefont{B.~E.} \bibnamefont{Feldman}},
  \bibinfo{author}{\bibfnamefont{I.}~\bibnamefont{Kimchi}},
  \bibinfo{author}{\bibfnamefont{A.~C.} \bibnamefont{Potter}},
  \bibinfo{author}{\bibfnamefont{Q.~D.} \bibnamefont{Gibson}},
  \bibinfo{author}{\bibfnamefont{R.~J.} \bibnamefont{Cava}},
  \bibinfo{author}{\bibfnamefont{A.}~\bibnamefont{Vishwanath}},
  \bibnamefont{and} \bibinfo{author}{\bibfnamefont{A.}~\bibnamefont{Yazdani}},
  \bibinfo{journal}{Nature Materials} \textbf{\bibinfo{volume}{13}},
  \bibinfo{pages}{851} (\bibinfo{year}{2014}).

\bibitem[{\citenamefont{Beidenkopf et~al.}(2011)\citenamefont{Beidenkopf,
  Roushan, Seo, Gorman, Drozdov, Hor, Cava, and Yazdani}}]{nphys2108}
\bibinfo{author}{\bibfnamefont{H.}~\bibnamefont{Beidenkopf}},
  \bibinfo{author}{\bibfnamefont{P.}~\bibnamefont{Roushan}},
  \bibinfo{author}{\bibfnamefont{J.}~\bibnamefont{Seo}},
  \bibinfo{author}{\bibfnamefont{L.}~\bibnamefont{Gorman}},
  \bibinfo{author}{\bibfnamefont{I.}~\bibnamefont{Drozdov}},
  \bibinfo{author}{\bibfnamefont{Y.~S.} \bibnamefont{Hor}},
  \bibinfo{author}{\bibfnamefont{R.~J.} \bibnamefont{Cava}}, \bibnamefont{and}
  \bibinfo{author}{\bibfnamefont{A.}~\bibnamefont{Yazdani}},
  \bibinfo{journal}{Nature Physics} \textbf{\bibinfo{volume}{7}},
  \bibinfo{pages}{939} (\bibinfo{year}{2011}).

\bibitem[{\citenamefont{Zeljkovic et~al.}(2014)\citenamefont{Zeljkovic, Okada,
  Huang, Sankar, Walkup, Zhou, Serbyn, Chou, Tsai, Lin et~al.}}]{nphys3012}
\bibinfo{author}{\bibfnamefont{I.}~\bibnamefont{Zeljkovic}},
  \bibinfo{author}{\bibfnamefont{Y.}~\bibnamefont{Okada}},
  \bibinfo{author}{\bibfnamefont{C.-Y.} \bibnamefont{Huang}},
  \bibinfo{author}{\bibfnamefont{R.}~\bibnamefont{Sankar}},
  \bibinfo{author}{\bibfnamefont{D.}~\bibnamefont{Walkup}},
  \bibinfo{author}{\bibfnamefont{W.}~\bibnamefont{Zhou}},
  \bibinfo{author}{\bibfnamefont{M.}~\bibnamefont{Serbyn}},
  \bibinfo{author}{\bibfnamefont{F.}~\bibnamefont{Chou}},
  \bibinfo{author}{\bibfnamefont{W.-F.} \bibnamefont{Tsai}},
  \bibinfo{author}{\bibfnamefont{H.}~\bibnamefont{Lin}}, \bibnamefont{et~al.},
  \bibinfo{journal}{Nature Physics} \textbf{\bibinfo{volume}{10}},
  \bibinfo{pages}{572} (\bibinfo{year}{2014}).

\bibitem[{\citenamefont{Lee et~al.}(2015)\citenamefont{Lee, Kim, Lee, Billinge,
  Zhong, Schneeloch, Liu, Valla, Tranquada, Gu et~al.}}]{pnas.1424322112}
\bibinfo{author}{\bibfnamefont{I.}~\bibnamefont{Lee}},
  \bibinfo{author}{\bibfnamefont{C.~K.} \bibnamefont{Kim}},
  \bibinfo{author}{\bibfnamefont{J.}~\bibnamefont{Lee}},
  \bibinfo{author}{\bibfnamefont{S.~J.~L.} \bibnamefont{Billinge}},
  \bibinfo{author}{\bibfnamefont{R.}~\bibnamefont{Zhong}},
  \bibinfo{author}{\bibfnamefont{J.~A.} \bibnamefont{Schneeloch}},
  \bibinfo{author}{\bibfnamefont{T.}~\bibnamefont{Liu}},
  \bibinfo{author}{\bibfnamefont{T.}~\bibnamefont{Valla}},
  \bibinfo{author}{\bibfnamefont{J.~M.} \bibnamefont{Tranquada}},
  \bibinfo{author}{\bibfnamefont{G.}~\bibnamefont{Gu}}, \bibnamefont{et~al.},
  \bibinfo{journal}{Proceedings of the National Academy of Sciences}
  \textbf{\bibinfo{volume}{112}}, \bibinfo{pages}{1316} (\bibinfo{year}{2015}).

\bibitem[{\citenamefont{Zhang et~al.}(2009)\citenamefont{Zhang, Cheng, Chen,
  Jia, Ma, He, Wang, Zhang, Dai, Fang et~al.}}]{PhysRevLett.103.266803}
\bibinfo{author}{\bibfnamefont{T.}~\bibnamefont{Zhang}},
  \bibinfo{author}{\bibfnamefont{P.}~\bibnamefont{Cheng}},
  \bibinfo{author}{\bibfnamefont{X.}~\bibnamefont{Chen}},
  \bibinfo{author}{\bibfnamefont{J.-F.} \bibnamefont{Jia}},
  \bibinfo{author}{\bibfnamefont{X.}~\bibnamefont{Ma}},
  \bibinfo{author}{\bibfnamefont{K.}~\bibnamefont{He}},
  \bibinfo{author}{\bibfnamefont{L.}~\bibnamefont{Wang}},
  \bibinfo{author}{\bibfnamefont{H.}~\bibnamefont{Zhang}},
  \bibinfo{author}{\bibfnamefont{X.}~\bibnamefont{Dai}},
  \bibinfo{author}{\bibfnamefont{Z.}~\bibnamefont{Fang}}, \bibnamefont{et~al.},
  \bibinfo{journal}{Phys. Rev. Lett.} \textbf{\bibinfo{volume}{103}},
  \bibinfo{pages}{266803} (\bibinfo{year}{2009}).

\bibitem[{\citenamefont{Okada et~al.}(2011)\citenamefont{Okada, Dhital, Zhou,
  Huemiller, Lin, Basak, Bansil, Huang, Ding, Wang
  et~al.}}]{PhysRevLett.106.206805}
\bibinfo{author}{\bibfnamefont{Y.}~\bibnamefont{Okada}},
  \bibinfo{author}{\bibfnamefont{C.}~\bibnamefont{Dhital}},
  \bibinfo{author}{\bibfnamefont{W.}~\bibnamefont{Zhou}},
  \bibinfo{author}{\bibfnamefont{E.~D.} \bibnamefont{Huemiller}},
  \bibinfo{author}{\bibfnamefont{H.}~\bibnamefont{Lin}},
  \bibinfo{author}{\bibfnamefont{S.}~\bibnamefont{Basak}},
  \bibinfo{author}{\bibfnamefont{A.}~\bibnamefont{Bansil}},
  \bibinfo{author}{\bibfnamefont{Y.-B.} \bibnamefont{Huang}},
  \bibinfo{author}{\bibfnamefont{H.}~\bibnamefont{Ding}},
  \bibinfo{author}{\bibfnamefont{Z.}~\bibnamefont{Wang}}, \bibnamefont{et~al.},
  \bibinfo{journal}{Phys. Rev. Lett.} \textbf{\bibinfo{volume}{106}},
  \bibinfo{pages}{206805} (\bibinfo{year}{2011}).

\bibitem[{\citenamefont{Kim et~al.}(2014)\citenamefont{Kim, Yoshizawa, Ishida,
  Eto, Segawa, Ando, Shin, and Komori}}]{PhysRevLett.112.136802}
\bibinfo{author}{\bibfnamefont{S.}~\bibnamefont{Kim}},
  \bibinfo{author}{\bibfnamefont{S.}~\bibnamefont{Yoshizawa}},
  \bibinfo{author}{\bibfnamefont{Y.}~\bibnamefont{Ishida}},
  \bibinfo{author}{\bibfnamefont{K.}~\bibnamefont{Eto}},
  \bibinfo{author}{\bibfnamefont{K.}~\bibnamefont{Segawa}},
  \bibinfo{author}{\bibfnamefont{Y.}~\bibnamefont{Ando}},
  \bibinfo{author}{\bibfnamefont{S.}~\bibnamefont{Shin}}, \bibnamefont{and}
  \bibinfo{author}{\bibfnamefont{F.}~\bibnamefont{Komori}},
  \bibinfo{journal}{Phys. Rev. Lett.} \textbf{\bibinfo{volume}{112}},
  \bibinfo{pages}{136802} (\bibinfo{year}{2014}).

\bibitem[{\citenamefont{Economou}(2006)}]{economou2006green}
\bibinfo{author}{\bibfnamefont{E.~N.} \bibnamefont{Economou}},
  \emph{\bibinfo{title}{Green's functions in quantum physics}},
  vol.~\bibinfo{volume}{7} (\bibinfo{publisher}{Springer Science \& Business
  Media}, \bibinfo{year}{2006}).

\bibitem[{\citenamefont{Biswas and Balatsky}(2010)}]{PhysRevB.81.233405}
\bibinfo{author}{\bibfnamefont{R.~R.} \bibnamefont{Biswas}} \bibnamefont{and}
  \bibinfo{author}{\bibfnamefont{A.~V.} \bibnamefont{Balatsky}},
  \bibinfo{journal}{Phys. Rev. B} \textbf{\bibinfo{volume}{81}},
  \bibinfo{pages}{233405} (\bibinfo{year}{2010}).

\bibitem[{\citenamefont{Istas et~al.}(2019)\citenamefont{Istas, Groth, and
  Waintal}}]{PhysRevResearch.1.033188}
\bibinfo{author}{\bibfnamefont{M.}~\bibnamefont{Istas}},
  \bibinfo{author}{\bibfnamefont{C.}~\bibnamefont{Groth}}, \bibnamefont{and}
  \bibinfo{author}{\bibfnamefont{X.}~\bibnamefont{Waintal}},
  \bibinfo{journal}{Phys. Rev. Res.} \textbf{\bibinfo{volume}{1}},
  \bibinfo{pages}{033188} (\bibinfo{year}{2019}).

\bibitem[{\citenamefont{Smidstrup et~al.}(2019)\citenamefont{Smidstrup,
  Markussen, Vancraeyveld, Wellendorff, Schneider, Gunst, Verstichel, Stradi,
  Khomyakov, Vej-Hansen et~al.}}]{SmidstrupJPCM2019}
\bibinfo{author}{\bibfnamefont{S.}~\bibnamefont{Smidstrup}},
  \bibinfo{author}{\bibfnamefont{T.}~\bibnamefont{Markussen}},
  \bibinfo{author}{\bibfnamefont{P.}~\bibnamefont{Vancraeyveld}},
  \bibinfo{author}{\bibfnamefont{J.}~\bibnamefont{Wellendorff}},
  \bibinfo{author}{\bibfnamefont{J.}~\bibnamefont{Schneider}},
  \bibinfo{author}{\bibfnamefont{T.}~\bibnamefont{Gunst}},
  \bibinfo{author}{\bibfnamefont{B.}~\bibnamefont{Verstichel}},
  \bibinfo{author}{\bibfnamefont{D.}~\bibnamefont{Stradi}},
  \bibinfo{author}{\bibfnamefont{P.~A.} \bibnamefont{Khomyakov}},
  \bibinfo{author}{\bibfnamefont{U.~G.} \bibnamefont{Vej-Hansen}},
  \bibnamefont{et~al.}, \bibinfo{journal}{Journal of Physics: Condensed Matter}
  \textbf{\bibinfo{volume}{32}}, \bibinfo{pages}{015901}
  (\bibinfo{year}{2019}).

\bibitem[{\citenamefont{González-Herrero
  et~al.}(2016)\citenamefont{González-Herrero, Gómez-Rodríguez, Mallet,
  Moaied, Palacios, Salgado, Ugeda, Veuillen, Yndurain, and
  Brihuega}}]{437.full}
\bibinfo{author}{\bibfnamefont{H.}~\bibnamefont{González-Herrero}},
  \bibinfo{author}{\bibfnamefont{J.~M.} \bibnamefont{Gómez-Rodríguez}},
  \bibinfo{author}{\bibfnamefont{P.}~\bibnamefont{Mallet}},
  \bibinfo{author}{\bibfnamefont{M.}~\bibnamefont{Moaied}},
  \bibinfo{author}{\bibfnamefont{J.~J.} \bibnamefont{Palacios}},
  \bibinfo{author}{\bibfnamefont{C.}~\bibnamefont{Salgado}},
  \bibinfo{author}{\bibfnamefont{M.~M.} \bibnamefont{Ugeda}},
  \bibinfo{author}{\bibfnamefont{J.-Y.} \bibnamefont{Veuillen}},
  \bibinfo{author}{\bibfnamefont{F.}~\bibnamefont{Yndurain}}, \bibnamefont{and}
  \bibinfo{author}{\bibfnamefont{I.}~\bibnamefont{Brihuega}},
  \bibinfo{journal}{Science} \textbf{\bibinfo{volume}{352}},
  \bibinfo{pages}{437} (\bibinfo{year}{2016}).

\bibitem[{\citenamefont{Hsieh et~al.}(2009{\natexlab{a}})\citenamefont{Hsieh,
  Xia, Qian, Wray, Dil, Meier, Osterwalder, Patthey, Checkelsky, Ong
  et~al.}}]{nature08234}
\bibinfo{author}{\bibfnamefont{D.}~\bibnamefont{Hsieh}},
  \bibinfo{author}{\bibfnamefont{Y.}~\bibnamefont{Xia}},
  \bibinfo{author}{\bibfnamefont{D.}~\bibnamefont{Qian}},
  \bibinfo{author}{\bibfnamefont{L.}~\bibnamefont{Wray}},
  \bibinfo{author}{\bibfnamefont{J.~H.} \bibnamefont{Dil}},
  \bibinfo{author}{\bibfnamefont{F.}~\bibnamefont{Meier}},
  \bibinfo{author}{\bibfnamefont{J.}~\bibnamefont{Osterwalder}},
  \bibinfo{author}{\bibfnamefont{L.}~\bibnamefont{Patthey}},
  \bibinfo{author}{\bibfnamefont{J.~G.} \bibnamefont{Checkelsky}},
  \bibinfo{author}{\bibfnamefont{N.~P.} \bibnamefont{Ong}},
  \bibnamefont{et~al.}, \bibinfo{journal}{Nature}
  \textbf{\bibinfo{volume}{460}}, \bibinfo{pages}{1101}
  (\bibinfo{year}{2009}{\natexlab{a}}).

\bibitem[{\citenamefont{Hsieh et~al.}(2009{\natexlab{b}})\citenamefont{Hsieh,
  Xia, Wray, Qian, Pal, Dil, Osterwalder, Meier, Bihlmayer, Kane
  et~al.}}]{science.1167733}
\bibinfo{author}{\bibfnamefont{D.}~\bibnamefont{Hsieh}},
  \bibinfo{author}{\bibfnamefont{Y.}~\bibnamefont{Xia}},
  \bibinfo{author}{\bibfnamefont{L.}~\bibnamefont{Wray}},
  \bibinfo{author}{\bibfnamefont{D.}~\bibnamefont{Qian}},
  \bibinfo{author}{\bibfnamefont{A.}~\bibnamefont{Pal}},
  \bibinfo{author}{\bibfnamefont{J.~H.} \bibnamefont{Dil}},
  \bibinfo{author}{\bibfnamefont{J.}~\bibnamefont{Osterwalder}},
  \bibinfo{author}{\bibfnamefont{F.}~\bibnamefont{Meier}},
  \bibinfo{author}{\bibfnamefont{G.}~\bibnamefont{Bihlmayer}},
  \bibinfo{author}{\bibfnamefont{C.~L.} \bibnamefont{Kane}},
  \bibnamefont{et~al.}, \bibinfo{journal}{Science}
  \textbf{\bibinfo{volume}{323}}, \bibinfo{pages}{919}
  (\bibinfo{year}{2009}{\natexlab{b}}).

\end{thebibliography}

\end{document}